\documentclass{article}                  %% LaTeX 2e
\usepackage{opex3}
\usepackage{graphicx}% Include figure files
\usepackage{dcolumn}% Align table columns on decimal point
\usepackage[mathlines]{lineno}
\usepackage{amsmath}
\usepackage{hyperref}
\usepackage{bm}% bold math
\usepackage{epstopdf}
\usepackage{epsfig}

\begin{document}

\title{Fiber propagation of vector modes}

\author{Bienvenu Ndagano$^1$, Robert Br\"{u}ning$^{1,2}$, Melanie McLaren$^1$, Michael~Duparr{\'e}$^2$ and Andrew Forbes$^{1,*}$}
\address{$^1$School of Physics, University of the Witwatersrand, Private Bag 3, Johannesburg 2050, South~Africa}
\address{$^2$ Institute of Applied Optics, Friedrich Schiller University Jena, Fröbelstieg 1, D-07743 Jena, Germany}
\email{$^{*}$andrew.forbes@wits.ac.za}

\begin{abstract}
Here we employ both dynamic and geometric phase control of light to produce radially modulated vector-vortex modes, the natural modes of optical fibers.  We then measure these modes using a vector modal decomposition set-up as well as a tomography measurement, the latter providing a degree of the non-separability of the vector states, akin to an entanglement measure for quantum states.  We demonstrate the versatility of the approach by creating the natural modes of a step-index fiber, which are known to exhibit strong mode coupling, and measure the modal cross-talk and non-separability decay during propagation.  Our approach will be useful in mode division multiplexing schemes for transport of classical and quantum states. 
\end{abstract}

 % NOTE: \ocis{} COULD/SHOULD BE ALIASED TO \pacs{} BUT
                          % MUST FORMAT THE TERMS CORRECTLY FOR OPEX

%\maketitle               % \maketitle IS A NULL FUNCTION in OPEX

%\begin{thebibliography}{}
%\begin{OEReferences}
%\begin{references}

%\end{thebibliography}

\section{ Introduction}
A key technique in achieving high-capacity optical communication is that of mode division multiplexing (MDM), which uses the spatial modes of light as information carriers \cite{li2014}, and has been shown promise in both free-space \cite{ wang2012, milione2015} and fiber \cite{ramachandran2013, richardson2013}. To date there has been work on MDM with orbital angular momentum (OAM) \cite{willner2015}, Bessel beams \cite{trichili2014,ahmed2014} and linearly polarized (LP) modes commonly used to describe fiber modes \cite{ryf2012}.  However, within free-space, atmospheric turbulence has been shown to negatively affect the strength of communication signals \cite{ibrahim2013}, and so MDM in fibers remains topical despite the challenges. 

The propagation of scalar spatial modes through fibers (the LP modes) has been limited to only a few modes, due to intermodal coupling within the fiber \cite{ramachandran2013}. This coupling is caused by inner fiber imperfections, such as index inhomogeneities, core ellipticity and eccentricity, as well as external perturbations like bends or strain \cite{Shemirani2009}. Despite these difficulties, successful fiber communication demonstrations have made use of specially designed fibers \cite{bozinovic2013, brunet2014}, and more recently using cylindrical vector modes \cite{ramachandran2009, gregg2015}. These modes are close to the natural modes of optical fibers as they consist of spatially inhomogeneous states of polarization \cite{Souza2007,Zhan2009}, but typically lack the radial modulation of the pure fiber modes.

In this paper, we exploit both the dynamic (optical path length) and geometric (Pancharatnam–Berry) phase changes of light to generate the natural modes of fibers: radial, azimuthal and polarization control.  We use a spatial light modulator to produce the radial modes before creating vector-vortex versions by geometric phase control \cite{dudley2013}. We introduce a composite diagnostic in the form of a vector mode decomposition that is mode specific\cite{milione2015}, as well as a new tomography approach that returns a quantitative measure of the degree of non-separability or ''vectorness'' of these modes \cite{mclaren2015}.  We use our set-up to create modes for a step-index fiber with known mode coupling \cite{flamm2013} and demonstrate that we are able to quantify the propagation of the natural modes of these fibers for the first time.  We quantify the effect of fiber perturbations on vector modes, which we demonstrate experimentally.  This approach will be useful in studies of MDM for increased bandwidth as well as in transport of high-dimensional quantum states down fibers.

\section{Concept and implementation}
\label{sec:setup}
\begin{figure}[h]
	\centerline{\includegraphics[width=12.0cm]{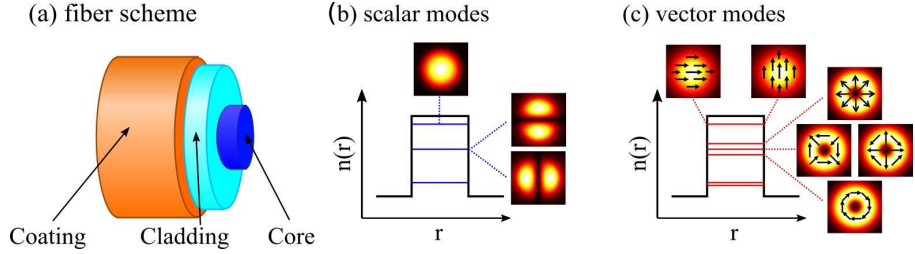}}
	\caption{Schematic of the step-index fiber and its modes; (a) structure of the fiber composed of a protective coating, cladding and core where the light is guided; schematic of  the mode patterns and spectrum for (b) the scalar approximation and (c) the natural vector modes with partialy resolved degeneracy of the first higher order mode group.}
	\label{fig:fiber}
\end{figure}

The full vector wave equation for a step-index fiber is given by \cite{snyder1983}
\begin{equation}
\{\bigtriangledown^{2}_{t} + n^{2}k^{2}\} \vec{e}_{t} + \bigtriangledown_{t} \{\vec{e}_{t} \cdot \bigtriangledown_{t} \left[\ln(n^{2})\right]\} = \beta^{2}\vec{e}_{t},
\label{eq:wave}
\end{equation} 
where $n$ is the refractive index of the fiber, $k=2\pi/\lambda$ is the wavevector, $\vec{e}_{t} $ is the transverse electric field and $\beta$ is the propagation constant of each vector mode solution. From Eq.~(\ref{eq:wave}), the first four higher-order vector solutions may be described as: transverse electric ($\textrm{TE}_{01}$), transverse magnetic ($\textrm{TM}_{01}$), hybrid electric odd ($\textrm{HE}^{\textrm{odd}}_{21}$) and even ($\textrm{HE}^{\textrm{even}}_{21}$). The two indices represent the number of half-wave patterns across the width and the height of the waveguide, respectively. The whole mode group is nearly degenerate, where the two $\textrm{HE}_{21}$ modes share the exact propagation constant, as shown in Fig.~\ref{fig:fiber}. As such, these modes are likely to couple during propagation within step-index fibers. 
Due to the cylindrical symmetry and step-like constant refractive index distribution of these fibers, the radial field distribution is given by Bessel-type functions:
\begin{equation}
E_{\ell p}(r) = 
\begin{cases}
J_{\lvert\ell\rvert}\left(\frac{u_{\ell p}r}{a}\right) / J_{\lvert\ell\rvert} \left(u_{\ell p}\right) & \text{for }~r < a\\
K_{\lvert\ell\rvert}\left(\frac{w_{\ell p}r}{a}\right) / K_{\lvert\ell\rvert} \left(w_{\ell p}\right)               & \text{for}~r \geq a
\end{cases} 
\label{eq:radial}
\end{equation}

Here, $a$ is the core radius, $J_{\ell}$ is the $\ell$th order Bessel function, $K_{\ell}$ is the $\ell$th order modified Bessel function with $u_{\ell p}$ and $w_{\ell p}$ are normalized propagation constants. In our experiment, we made use of a 30 mm long step-index fiber with a core radius $a = 15~ \mu$m and a numerical aperture $\textrm{NA} = 0.08$. For ease of alignment, we used a wavelength of 633 nm resulting in a total of 76 possible modes that can exist in the fiber.

\begin{figure}[h]
	\centerline{\includegraphics[width=12.0cm]{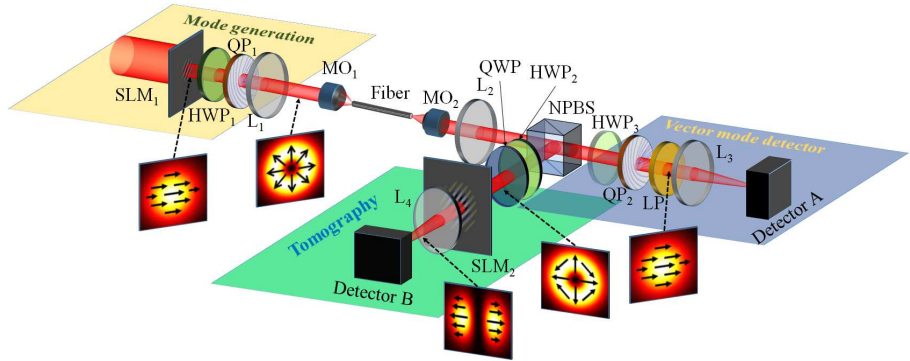}}
	\caption{Experimental setup used to generate and detect vector beams after propagating through a fiber by two separate measurement techniques: vector mode detection and tomography. Insets depict beam profiles at different planes. SLM - spatial light modulator; HWP - half wave plate; QP - q-plate; MO - microscopic objective; L - lens; NPBS - non-polarizing beam splitter; LP - linear polarizer; QWP - quarter wave plate. }
	\label{fig:setup}
\end{figure}

Owing to the radial distribution of the fiber, it is important to match the input field to the corresponding fiber mode group. As such, we exploit both the dynamic and geometric phase to generate the natural vector modes of the fiber, with our experimental setup shown in Fig.~\ref{fig:setup}. 
The first component of the experiment generated the natural fiber modes by radial, azimuthal and polarization control, labeled \textit{mode generation} in Fig.~\ref{fig:setup}. Firstly, the fundamental fiber mode, described by Eq.~(\ref{eq:radial}) with $\ell = 0$ and $p = 1$, was encoded onto a spatial light modulator (SLM) by complex amplitude modulation \cite{arrizon2007}. Secondly, we made use of a q-plate \cite{marrucci2006,li2012} to prepare vector vortex modes. Polarization is coupled to OAM by the selection rules: $\left| \ell, L \right\rangle \rightarrow \left| \ell + Q, R \right\rangle$ and $\left| \ell, R \right\rangle \rightarrow \left| \ell - Q, L \right\rangle$. The azimuthal charge introduced by the q-plate is $Q = 2q$. In our experiment $q = 1/2$ and the incident beam generated on SLM 1 was horizontally polarized with $\ell = 0$, which generated a TE$_{01}$ or azimuthally polarized vortex mode consisting of a superposition of $\ell = \pm 1$ modes. 
Thus, the field after the generation step can be described by
\begin{equation}
 \left| \Psi \right\rangle = \mbox{TE}_{01}(r,\phi) = E_{11}(r) \left(\exp(i\phi) \left| R \right\rangle - \exp(-i \phi) \left| L \right\rangle \right),
\label{equ:wave}
\end{equation}
where the field singularity introduced by the q-plate, results in a variation in the radial component $E_{\ell p}$ as well as in the azimuthal ($\phi$) degree of freedom. Note the OAM and polarization are not separable in Eq.~\ref{eq:wave}, and as such  $\Psi$ is reminiscent of a two-dimensional maximally entangled quantum state (qubit), a fact that we will exploit in our detection of the modes after the fiber.   

After propagation through the fiber the output modes were directed to the detection scheme comprising of two parts: a vector mode detection and a tomography measurement. The former acts as a vector mode decomposition technique for the fiber mode group \cite{milione2015}, which we implemented for the first four higher-order fiber modes.  For simplicity, we describe the single channel as shown in Fig.~\ref{fig:setup}, however in reality we implemeted a four-channel detection for each of the fiber modes, which were measured simultaneously. The output fiber mode was directed through a half-wave plate oriented at 0 rad followed by a q-plate ($q = -1/2$); it can be shown that this combination forms an inner product measurement in an analogous manner to that used for scalar mode detection \cite{flamm2013}. This pairing was followed by a linear polarizer, oriented to transmit horizontally polarized light. The final component of the decomposition measured the on-axis intensity on a camera after a Fourier lens $\textrm{L}_{3}$ \cite{flamm2013}. When the input mode matches the filter for the detected mode, a bright on-axis intensity is observed, otherwise zero on-axis intensity is measured. This detection scheme therefore determines whether the fiber causes any intermodal coupling between the four first higher order fiber modes. 

The second analysis technique, the \textit{tomographic measurement} borrows techniques from traditionally quantum measurements \cite{Jack2009} and is used to determine the effect of the fiber on the degree of non-separability or vectorness of the output mode. As non-separablity is not unique to quantum mechanics, many of the tools used to measure quantum states are applicable to vector beams \cite{mclaren2015}. We now apply this tool for the analysis of the natural modes in fibers. Projective measures are first performed on the polarization state, while the OAM degree of freedom was measured using holograms encoded onto SLM 2. As SLMs are polarization sensitive, a linear polarizer could not be used to measure the polarization states, as is commonly performed. Instead, a half-wave plate was inserted before SLM 2, and rotated to specific orientations to realize a filter for the linear polarization states: horizontal, vertical, diagonal and anti-diagonal. By inserting a quarter-wave plate, we were able to filter the two circularly polarized components, resulting in a total of six polarization measurements. Similarly, we created six different holograms on SLM 2 to represent the two pure OAM modes as well as four orientations of the superposition states: $\left|\ell~=~1~\right\rangle~+~\exp(i\theta_{2})\left|\ell~=~-1~\right\rangle$, for $\theta_{2} = 0, \pi/2, \pi, 3\pi/2$. A modal decomposition was performed for each polarization state by measuring the on-axis intensity on a camera situated after a Fourier lens $\textrm{L}_{4}$. This tomographic method produces an over-complete set of 36 measurements, which can be used to minimize the Chi-square quantity and reconstruct the density matrix $\rho= \left|\psi\right\rangle \left\langle\psi \right|$ of the state of the vector mode (e.g., that of Eq.~(\ref{equ:wave})).  This approach is analogous to measurements of qubit states except here one degree of freedom is polarization and the other spatial.  The degree of non-separability of any field can then be calculated from the density matrix. The concurrence of a state typically represents the degree of entanglement in a quantum entanglement experiment \cite{wootters2001}. However, as the vector modes are non-separable in the polarization and OAM degrees of freedom, the concurrence provides a measure of the degree of non-separability of the vector mode.
%It is given by \cite{wootters2001}
%\begin{equation}
%	\mathcal{C}(\rho) = \max\{0,\sqrt{\lambda_1} - \sqrt{\lambda_2} - \sqrt{\lambda_3} - \sqrt{\lambda_4}\} ,
%\end{equation}
%with $\lambda_i$ being the eigenvalues in decreasing order of the Hermitian matrix $\sqrt{\sqrt{\rho} \tilde{\rho} \sqrt{\rho}}$, where $\tilde{\rho}$ is the spin-flipped state of the density matrix $\rho$. As such, a measure of 1 represents a completely vector mode that is non-separable, while a measure of 0 represents a purely scalar mode, where the polarization and spatial intensity pattern of the mode can be separated.

\section{Experimental results and discussion}
\begin{figure}[h]
	\centerline{\includegraphics[width=11.5cm]{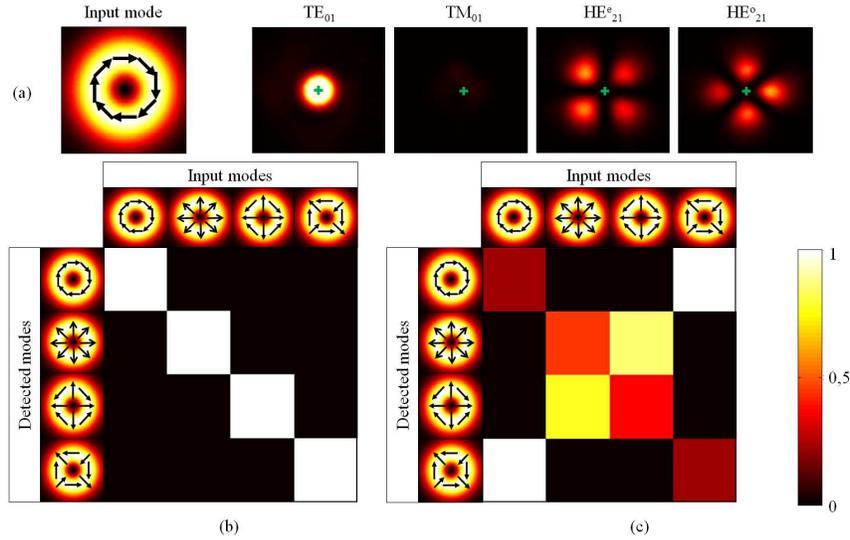}}
	\caption{Example and results of the vector mode decomposition; (a) depicts the working principle by illumination with a TE$_{01}$ mode. The bright spot in the TE channel on the optical axis, indicated by the green crosses, compared to zero intensity in the other channel denotes the high mode purity.  Illustration of the modal transmission matrix through (b) free space, indicating the high quality of the vector beam generation, and (c) after fiber transmission, showing the appearance of mode coupling during propagation through the fiber.}
	\label{fig:vector}
\end{figure}

Using the vector mode detection techniques described above, the results demonstrate intermodal coupling within the fiber between the first higher fiber modes (as expected). Figure~\ref{fig:vector}(a) shows an example of the vector decomposition results, where a $\textrm{TE}_{01}$ mode was generated after the q-plate and the on-axis intensity was measured in the four channels. It is clear from the images that a signal was only detected in channel $\textrm{TE}_{01}$. The comparison of the decomposition performed before and after the fiber, as shown in Fig.~\ref{fig:vector}(b) and (c), respectively, illustrates the intermodal coupling between the four modes within the fiber.  

\begin{figure}[h]
	\centerline{\includegraphics[width=12.0cm]{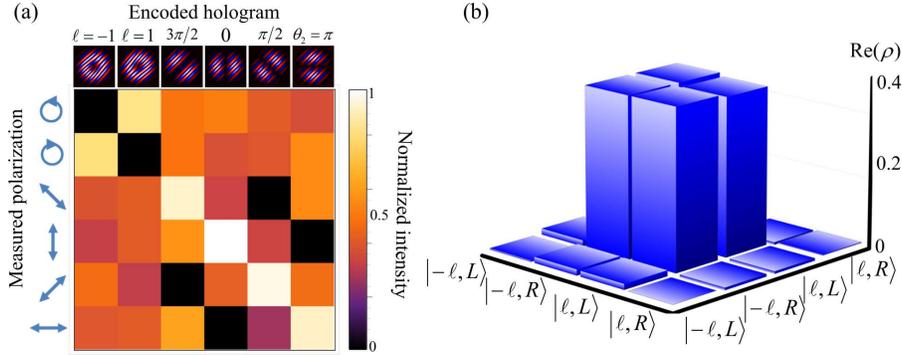}}
	\caption{Experimental results obtained from the full state tomography of the TE$_{01}$ mode, for qualitative analysis by rotating polarizer (\textcolor{blue}{Media˜1}). (a) Tomography matrix obtained by projection of the mode into six polarization and the corresponding six OAM states, measurement procedure (\textcolor{blue}{Media˜2}). (b) The qubit vector density matrix showing the high degree of non-separability of the investigated mode. }
	\label{fig: tomography matrix}
\end{figure}
From the 36 tomographic measurements shown in Fig.~\ref{fig: tomography matrix}(a), the qubit vector density matrix was calculated as shown in Fig.~\ref{fig: tomography matrix}(b). We found the concurrence, a measure of the degree of non-separability, of our vectorial field to be $ 0.98 \pm 0.01$. A maximal vector field is represented by 1, while a value of 0 represents a purely scalar field. Thus, although our first measurement scheme reported intermodal coupling within the fiber, the output mode has maintained a high degree of non-separability. We also measured the degree of non-separability for a higher order radial mode group with $\ell = 0$ and $p = 2$ in Eq.~(\ref{eq:radial}). We found it to be lower than that of the first mode group with a value of $0.85 \pm 0.01$. Since this mode group shows a lower power fraction confinement within the core than the first mode group \cite{snyder1983}, this mode group was more strongly affected by fiber imperfections, thereby resulting in a stronger decay of non-separability.

This technique enables the investigation of the effects of fiber perturbations on the degree of non-separability of the vector modes. Table~\ref{table:concurrence} shows the effect of adding stress to a fiber. By stressing the fiber, the energy exchanged between the modes fluctuates in a sinusoidal manner. Depending on the level of stress, the coupling between the modes can be increased or decreased. As the coupling length is inversely proportional to the coupling strength, modes that couple weakly will have longer coupling lengths. Thus, increasing the stress on the fiber does not directly result in a decrease in the degree of non-separability and as such we measure a degree of non-separability of 0.43 and 0.73 for stresses of 0.5 N and 1 N, respectively.
\begin{table}[h]
	\caption{Effects of stress on degree of non-separability} % title of Table
	\centering % used for centering table
	\begin{tabular}{c c } % centered columns (4 columns)
		\hline\hline %inserts double horizontal lines
		Stress (N) & Degree of non-separability \\ [0.5ex] % inserts table
		%heading
		\hline % inserts single horizontal line
		0 & 0.98 $\pm$ 0.01  \\ % inserting body of the table
		0.5 & 0.43 $\pm$ 0.01  \\
		1.0 & 0.73 $\pm$ 0.01 \\ [1ex] % [1ex] adds vertical space
		\hline %inserts single line
	\end{tabular}
	\label{table:concurrence} % is used to refer this table in the text
\end{table}

\section{Conclusion}
Here we have demonstrated a complete generation and detection scheme for the natural modes in fibers.  We use a spatial light modulator to create radially modulated modes through complex amplitude modulation, and then exploit geometric phase change to control the polarization and OAM of the modes.  We have demonstrated a composite detection method with which to analyze the effects of fiber propagation on these modes, using step-index fibers as an example. This technique can be extended to investigate the fiber propagation of unknown vector beams \cite{flamm2010}. As the non-separability of vector modes offer increased potential for optical communication in both the classical and quantum regimes and as such, the ability to quantify the degree of non-separability is a useful technique, and the first time it has been applied to fiber mode analysis. 

\section*{Acknowledgements}
We would like to thank Michael Escuti for the fabrication of the q-plates, and the National Research Foundation of South Africa as well as the African Laser Centre for financial support. 

\end{document}